**Title: When the atoms dance: exploring mechanisms of electron-beam induced modifications of materials with machine-learning assisted high temporal resolution electron microscopy**

**Authors:** Matthew G. Boebinger[1*], Ayana Ghosh[2], Kevin M. Roccapriore[1], Sudhajit Misra[1], Kai Xiao[1], Stephen Jesse[1], Maxim Ziatdinov[1,2], Sergei V. Kalinin[3], Raymond R. Unocic[1*]

[1] Center for Nanophase Materials Sciences, Oak Ridge National Laboratory, Oak Ridge, TN, USA

[2] Computational Science and Engineering Division, Oak Ridge National Laboratory, Oak Ridge, TN, USA

[3] Department of Materials Science and Engineering, University of Tennessee, Knoxville, TN, USA

**Abstract:**

Directed atomic fabrication using an aberration-corrected scanning transmission electron microscope (STEM) opens new pathways for atomic engineering of functional materials. In this approach, the electron beam is used to actively alter the atomic structure through electron beam induced irradiation processes. One of the impediments that has limited widespread use thus far has been the ability to understand the fundamental mechanisms of atomic transformation pathways at high spatiotemporal resolution. Here, we develop a workflow for obtaining and analyzing high-speed spiral scan STEM data, up to 100 fps, to track the atomic fabrication process during nanopore milling in monolayer $MoS_2$. An automated feedback-controlled electron beam positioning system combined with deep convolution neural network (DCNN) was used to decipher fast but low signal-to-noise datasets and classify time-resolved atom positions and nature of their evolving atomic defect configurations. Through this automated decoding, the initial atomic disordering and reordering processes leading to nanopore formation was able to be studied across various timescales. Using these experimental workflows a greater degree of speed and information can be extracted from small datasets without compromising spatial resolution. This approach can be adapted to other 2D materials systems to gain further insights into the defect formation necessary to inform future automated fabrication techniques utilizing the STEM electron beam.

**Introduction:**

The rapidly evolving landscape of science and technology demands unprecedented precision in materials fabrication, pushing the boundaries of traditional manufacturing techniques. Atomically precise fabrication has emerged as a transformative approach that promises unparalleled control over the structure and properties of materials at the atomic scale. This precision is essential for unlocking breakthroughs in diverse fields, including quantum sensing and communications [1-4]. Recently, it has been shown that the electron beam can be used as an exceptionally precise tool for direct fabrication on the atomic scale, the approach dubbed the atomic forge [5]. As the pursuit of atom-by-atom construction of materials gains momentum [2, 6-16], it becomes imperative to deepen our understanding of the underlying processes involved in atomic disordering and reordering. This knowledge is crucial in informing the development of future atomic fabrication techniques for novel functional applications. By comprehending and identifying the mechanisms behind defect formation, it becomes possible to tailor defects within two-dimensional (2D) materials to exhibit specific electronic, magnetic, or optical properties [17-26]. Such tailored defects hold the potential to revolutionize the performance and functionality of 2D materials, opening exciting avenues for scientific exploration and technological innovation in diverse fields. In this paper, we aim to explore the intricate relationship between defect formation processes and electron beam irradiation, offering insights into the design principles for atomically precise fabrication and the creation of functional materials with tailored properties.

While traditional synthesis methods, such as top-down or bottom-up synthesis have been used extensively for 2D materials [27, 28], these methods do not reach the needed length scales required for atomic structure fabrication. With the advent of using the sub-Å $e^-$ beam of a scanning transmission electron microscope (STEM) as a direct fabrication tool in recent years, information has already been gathered on the effect that the $e^-$ beam irradiation has on various 2D materials [11-16, 29-33]. This technique relies on the localized defect generation caused by the highly energetic electrons. The primary mechanism for electron beam damage is believed to be the knock-on effect, where the electron transfers enough kinetic energy to the nucleus to displace it from the atomic structure [33-36]. The knock-on energy is determined by the momentum transfer from the electron to the nucleus and the binding energy of the atom in the lattice. However, in most materials, atomic manipulation can be observed well below this knock-on threshold, stimulating the development of alternative theories for this process. In previous work, Susi proposed the additional role of the phonons to the displacement [33]. At the same time, Lingerfelt and Meunier proposed and developed theories based on the local electronic excitations to the antibonding states and subsequent isomerization of the excited bond, akin to the photochemical effects [37, 38]. Additionally, recently TMD semiconductors/semimetals have been seen to suffer from beam induced charging. At slower scan speeds sample charging can modify the effective knock-on threshold energy. Regardless, by controlling these atomic interactions, different defects and structures can be written into the atomic structure of 2D materials [11-16, 29-33]. Through precise control with the $e^-$ beam these defects can be fabricated while limiting the overall damage to the surrounding atomic structure adding more functionality to possible devices made with 2D materials.

Here we used the transition metal dichalcogenide (TMD) $MoS_2$, as the model system to explore the structural changes associated with defect and nanopore formation. As a model system, defect engineering within TMDs have been of interest in recent years as different atomic defects [39-42], nanopores [17, 24, 43], and nanowire structures [16, 21, 44-46] fabricated within the monolayer have been shown to have beneficial for electronic and optical properties. Using an external beam control system operating at high temporal resolution, the atomic reordering and disordering that initially takes place during $e^-$ beam irradiation was recorded. These images were then decoded using the ELIT workflow [47, 48] to process the images and display atomic trajectories and the formation of different defect structures. This workflow allowed for more information to be extracted from small datasets without losing out on temporal or spatial resolution.

To acquire the atomically resolved images required for these experiments, an aberration-corrected STEM was used in conjunction with an external beam control system that allows for simultaneous imaging and atomic fabrication. A custom-built beam control system was used for these studies due to the difficulties introduced from traditional STEM raster scanning [49-51]. The standard usage of these methods results in high quality atomic resolution data, but at the expense of slower scanning rates that result in higher $e^-$ beam irradiation to the samples and less frequent updates on the beam-induced modifications. However, when using the STEM $e^-$ beam as the fabrication tool, greater control at higher temporal resolution is needed for automated feedback control. Therefore an external beam control system, that was developed at ORNL, was used to control the scanning pathways the beam takes while simultaneously imaging [6, 8, 31, 49-51]. Through this novel beam control system, a spiral scan path is used where single frequency drive signals to the scan coils is employed, enabling acquisition of atomically resolved images at speeds of up to 100 frames per second (fps) [51]. The utilization of a spiral path eliminates sharp changes in scan velocity from one line to the next and mitigates the flyback duration associated with traditional raster scan profiles, enabling much faster imaging speeds without distortion, improved temporal resolution from frame to frame, and therefore reduced electron dose to the material per frame while using standard beam steering magnetic coils. At the periphery the beam is moving faster likely causing less charge build-up localizing the beam irradiation effects to the center of the FOV frames. Remaining image distortion resulting from high-speed imaging can be mitigated using subsequent image correction methods [50, 51]. However, at such high imaging speeds, averaging times per pixel are lower and result in lower signal-to-noise ratios in each frame. Here we show that through the development and application of advanced post-processing techniques and sophisticated data analytics, we are able to extract critical information from the images (specifically, atomic positions and species) that can be used to better understand beam-induced material transformations as they occur. Furthermore, the beam control system is built with automation and feedback that is used to assess beam damage from frame to frame and automatically stop imaging after a specified damage threshold has been crossed (discussed further below).

In this study, a workflow incorporating deep learning (DL) networks was employed to decode individual frames captured during this rapid image acquisition. Distinguishing features in atomically resolved experiments, such as scanning tunneling microscopy (STM) and scanning transmission electron microscopy (STEM), differ from classifying objects commonly found in publicly available datasets like CIFAR and ImageNet [52, 53]. While previous DL networks have been designed and implemented for accurate image analysis of molecular [54] and atomically resolved data, [55-59] they typically require high-quality data for precise model predictions. In this work, a network was utilized for image segmentation of nearly identical objects, namely atoms, while also quantifying prediction uncertainty to adapt to images originating from different experimental or simulation setups. Analysis of the acquired datasets presented challenges due to material lattice distortions, image drifts, and variations in imaging parameters, resulting in an out-of-drift distribution effect that reduces the accuracy of traditional deep convolutional neural network (DCNN) model predictions. To address this issue, the ensemble learning-iterative training (ELIT) workflow proposed and developed by Ghosh et al. was implemented here [47, 48]. This approach demonstrated how a DL workflow could be applied to STEM data, allowing for adaptation to changes in imaging conditions. Within the ELIT workflow, multiple U-Net models were trained with random initialization parameters. The ensemble training phase aimed to select artifact-free models for feature selection, while the iterative phase directed the network's attention towards recognized features to identify unknowns while considering pixel-wise uncertainty. By following this process, the models successfully identified the location and type of each atom within the frame with a high degree of certainty. Through the utilization of the ELIT workflow, these small datasets were effectively decoded, uncovering new insights into various materials systems while simultaneously increasing the speed and information content.

**Results and Discussion:**

The experimental workflow from data acquisition to processing and decoding can be seen in Figure 1. The process starts with the creation of ideal simulated training to give a baseline for the models constructed later to iterate from (Figure 1a). To begin crystal models of pristine molybdenum disulfide's atomic structure, with a random assortment of sulfur sites deleted, were used as training data. These sulfur site point defects were introduced into the training data to simulate the defect generation that occurs during the e$^-$ beam fabrication therefore it was deemed necessary to train the model on the distinction of sites with both two overlapping sulfur atoms, as is seen in the ideal crystal structure, and sites with only a single sulfur atom. STEM images were then simulated using the variation of the multislice algorithm within the abTEM package developed by Madsen et al. [60, 61] Using these simulated images, corresponding circular masks with a fixed radius were constructed classifying each atomic site appropriately. Once created, these image and mask pairs were augmented through random cropping, rotations, zooming-in, contrast variations and through the application of blurring, Gaussian and Poisson noise to replicate real experimental data more accurately. An example of a final simulated image mask pair can be seen in the block inset of Figure 1a.

The experimental data was acquired using an aberration corrected Nion UltraSTEM100 operating at 60kV with a spatial resolution of ~1Å. The custom-built scanning feedback and electron beam control system was controlled using LabView software to acquire atomically resolved image datasets with the high angle annular dark field (HAADF) detector. A schematic of this system's operation can be seen in Figure 1c. The spiral beam pathway discussed previously was used for these experiments to maximize the temporal resolution without sacrificing spatial resolution. Prior to each fabrication step, predetermined parameters were set for the maximum number of spirals (i.e. the maximum e$^-$ irradiation dosage) as well as the spiral size and speed. These spirals are made to maintain content frequency drive signal to the scan coils to avoid frequency dependance. The dose profile can be controlled to a certain degree, but by default there is less dose in the periphery as there is a lower density of the scan path points at the edge of the FOV. For these experiments the spiral field of view was maintained at around 1.5 to 2.5 nm in diameter for atomic resolution, with a temporal resolution ranging from 5 to 100 fps (200-10 milliseconds). During the fabrication process, the beam control system acquired an image after each scan and plotted the average intensity of each frame, which, as the formation of defects began within the MoS$_2$ monolayer, lowered as the process continued, supporting the notion of material ejection. After reaching a pre-determined intensity threshold, the fabrication process was stopped, and the experimental datasets were exported to study the atomic disordering that took place prior to nanopore formation. Due to the need to increase this system's speed, the exported data results in few pixels per frame, therefore pre-processing of this data was needed prior to decoding through the DCNN models. The data was first resized, and images were cropped appropriately to match the square image training data pixel to angstrom ratio before applying similar blurring and noise application as was previously used for the training data.

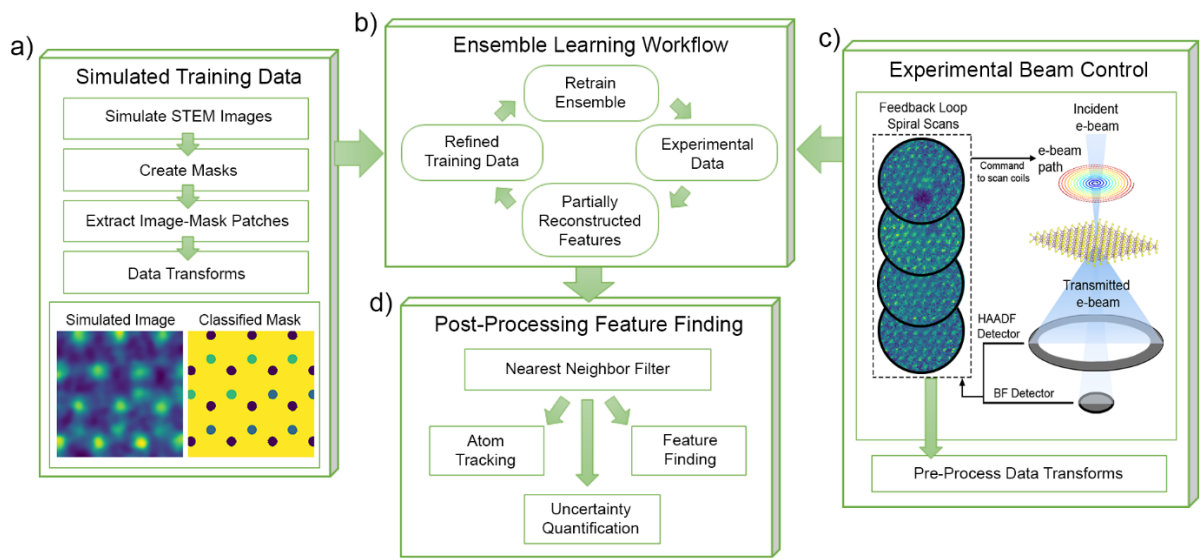

**Figure 1:** Block diagram of the steps in the experimental workflow: **(a)** Simulate the training data: The initial block consisted of constructing the simulated training data. First STEM images are simulated and then their associated labeled mask pairs created, followed by data transforms. **(b)** The Ensemble Learning and Iterative Training workflow is then demonstrated as the training and experimental data is input and several DCNN models are output comparing the fit between the model constructed off the simulated data vs. the experimental. These models are then compared against one another until an ideal model is selected. **(c)** A schematic of the experimental automated feedback-controlled beam control experiment that allows for *in situ* monitoring and study of the atomic disordering of MoS$_2$ using high temporal resolution. **(d)** The output of the models from the ELIT workflow is then fed through post-processing filters before feature finding can be performed.

With both simulated and experimental data acquired, the ensemble learning and iterative training workflow (ELIT) workflow, as demonstrated by Ghosh et al. [47], was used to identify features from the low resolution data using these datasets as the initial training set and inputs respectively (Figure 1b). This approach is known to result in the overdetermination of the atomic sites, to avoid this problem several post-processing steps were taken before features could be extracted. The output from these models unrealistically oversampled the certain atomic sites on several frames, labeling overlapping atoms on single sites. A nearest neighbor filter was created and used to correct this issue as the last step in Figure 1d. This filter consisted of three major steps: a nearest neighbor distance filter, using atom finding to locate the correct sites, and finally a filter refinement. The first step consisted of a nearest neighbor distance filter that found the clusters of problem sites by setting a distance threshold of ~0.85 Å, which is about half the distance between the nearest atoms to avoid single sites being identified as different classes multiple times. This was followed by implementing atom finding through the STEMtool package developed by Mukherjee et al. [62] on the decoded images for each class, being Mo sites (red points), projected S sites with two atoms (S2, yellow points), and projected S sites with only one S atom (S1, green points), to find the correct center of mass for each site. After these coordinates were found, they were cross-correlated with the clusters of decoded coordinates found from the first step, with the mislabeled sites being deleted. This left us with a fully decoded image and associated coordinate list for each frame within the dataset. This decoded data could then be used for feature finding.

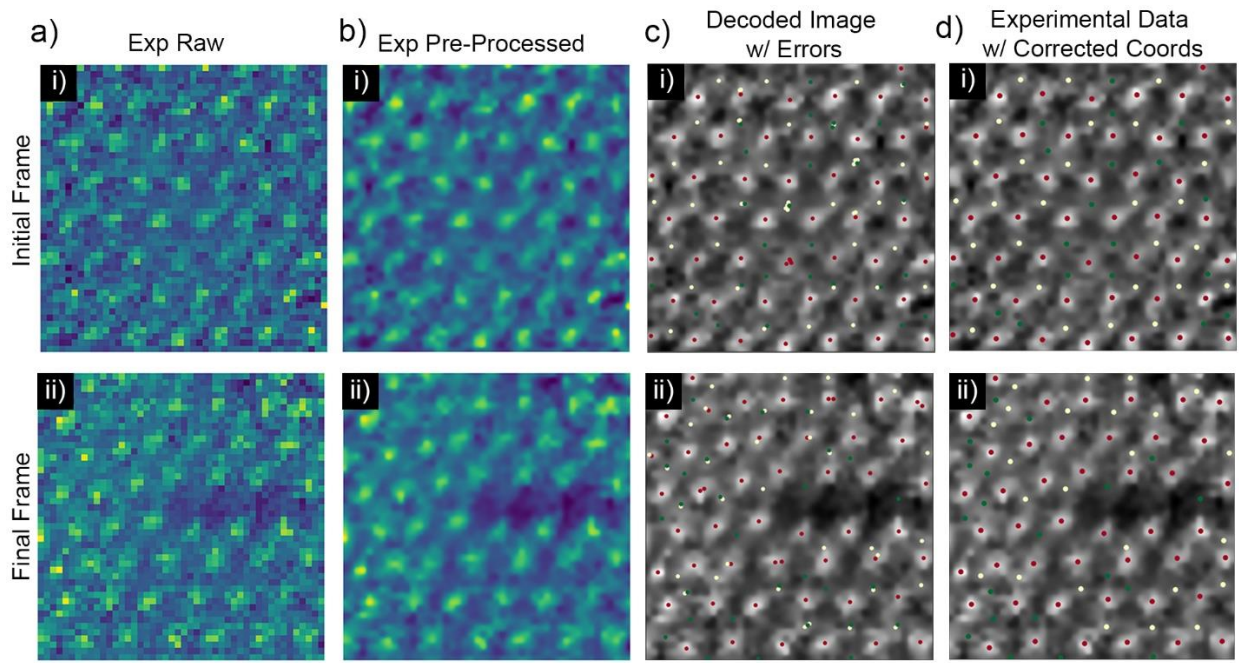

**Figure 2:** Initial (i) and Final (ii) frames taken of the atomic disordering shown at different stages of decoding. **(a)** The raw experimental image data before any processing. **(b)** Pre-processing of the experimental data was performed by first denoising the data and then using a gaussian filter. **(c)** The decoded results of the ELIT workflow are plotted over the gaussian-filtered experimental frame. As seen, there are several sites that are labeled incorrectly, resulting in clusters of atoms that are not realistic. **(d)** The results of the nearest neighbor filter with the redundant points and incorrect labeling removed from the frame. Every frame shown has the same field of view of 2.56x2.56 nm.

Due to the effectiveness of the ELIT workflow in decoding the experimental data, the temporal resolution was able to be pushed to much higher limits using this beam control system while still being capable of gathering structural information from the atomic disordering. Frames taken at different temporal resolutions and their associated decoded frames are shown in Figure 3. Scan rates varying from 10 to 100 fps were performed using the external beam control system. Frames were then successfully decoded as seen in the ii and v frames, with red representing the Mo atoms, green the S2 sites, and blue the S1 sites. The processed data with the corrected coordinates can also be seen with the same color distinctions as used previously. The pre-processing and decoding begin to breakdown at 100 fps as the bright Mo are the only sites that can be reliably identified due to lack of current for each scan. However even at 50 fps, with each frame being acquired every 20 ms, atomic resolution is still fully achievable, and the S sites can be correctly identified as containing a $V_S$ (S1) or two S atoms (S2).

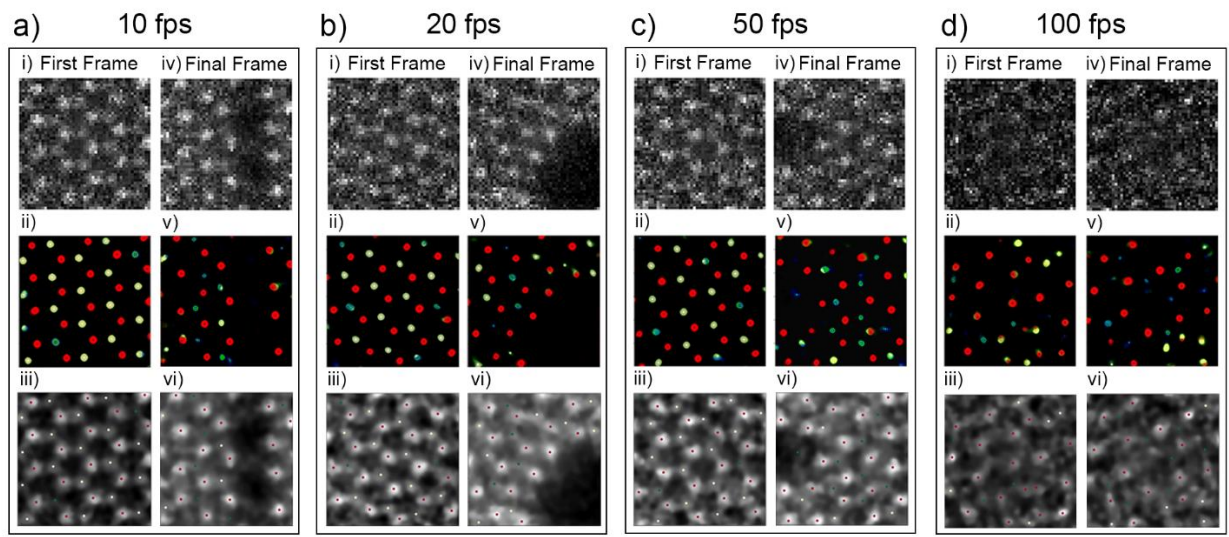

**Figure 3: (a-d)** Initial (i) and final (iv) frames of taken at different scan rates at 10, 20, 50 and 100 fps. These slides were then decoded into the Mo (red), S2 (green), and S1 (blue) classes in (ii) and (v) respectively. The corrected coordinates are overlaid onto the processed frames (iii) and (vi) respectively. Every frame has the same field of view of 1.6x1.6 nm.

With the image datasets fully decoded into the possible atomic sites, information can be obtained as to the stoichiometry, structure and atomic movement that takes place prior to the nanopore formation. By retaining the classifier information, the exact number of Mo and S atoms within the field of view being irradiated can be tracked and recorded along with the associated dose. Using these decoded frames the average dosage (assuming an average beam current of around 50 pA for the day) required to cause initial atomic disordering as well as form sulfur single vacancy line (SVL) defects and point defects ($9.0 \pm 4.5 \times 10^8\ e^-/nm^2$) that act as the nucleation site for expanding nanopores can be tracked for each experimental milling process. On average, we find that initial disordering takes a dose of $3.3 \pm 2.3 \times 10^8\ e^-/nm^2$ to begin. After initial deformation, an additional $1.3 \times 10^8\ e^-/nm^2$ is needed to form SVLs, when they did form ~75% of the time. From disordering to formation of a vacancy point defect, or hole, an additional dose of $5.7 \times 10^8\ e^-/nm^2$ is applied to the MoS$_2$ field of view. The dosage required to induce atomic transformations was found to generally be dependent on the speed of the scanning parameters with the slower scans with longer dwell times inducing atomic transformations more rapidly. However, the dosage required for each respective transformation was found to be irregular even within similar settings indicating other parameters outside of only dosage influencing the formation of these defects. Additionally, trends in the location of the initial disordering and defect can be studied in each milling dataset. As expected, the nucleation of the nanopore is entirely dependent on the location of the atomic disordering with the center of the frame requiring on average, lower dosage to begin forming in the center of the frame. When the initial defects begin forming at the edge of the frame, they take much longer to grow due to lower local dosage. It was also found that smaller fields of view result in the disordering further from the center of the field of the view.

Outside of general dosage trend, atomic tracking can be conducted between frames. Tracking is done by finding the closest identified atom between frames within a set distance threshold, if this distance is overcome, the atom has either been ejected or completely displaced. However, this threshold is large enough to observe the atomic shift both Mo and S species exhibit during disordering. Several examples of the spatiotemporal trajectories of single atomic sites at different acquisition rates were gathered in a brush diagram in Figure 4 allowing for the atomic disordering of both Mo (red) and S (blue for S2 site and green for S1 site or a single S vacancy) to be clearly observed. By tracking the classifier of the atom sites the formation of single and double $V_S$ is clearly shown as well.

In Figure 4a, the 5 fps experiment demonstrates the clear switch between S2 and S1 sites throughout the milling process. As a split in the lattice forms (Fig. 4a(iii)) prior to the opening of the nanopore the tracking become erratic and the distance between selected neighboring Mo and S sites increases before the S site disappears in some frames. Figure 4b shows a smaller field of view at a faster acquisition of 10 fps and clearly shows the formation of sulfur single vacancy line (SVL) defects. In the final frame in (iii) several of these SVL defects converge together before a nanopore begins to form. In the tracking plot one can clearly see when the sulfur vacancy, that makes up one SVL forms, which is quickly followed by the Mo and S site to move closer together as SVLs cause a distortion as the lattice compresses. In the 20 fps experiment in Figure 4c, the nanopore forms early in the top left corner of the field of view. As the nanopore grows, the Mo and S atom tracking can be seen to slowly shift as the lattice is compressed while the nanopore grows. Additionally, the S site remains S2 even after extended $e^-$ beam irradiation, this was likely due to the formation of the nanopore. Finally in the 50 fps experiment in Figure 4d, the more random movement of the tracking column associated with higher temporal resolution better captures erratic nature of atomic motion within crystal structures while being exposed to $e^-$ beam irradiation. All of these examples clearly emphasize that atomic structural information can be acquired even at 50 fps without relying upon highly specialized equipment or modifications to the electron optics.

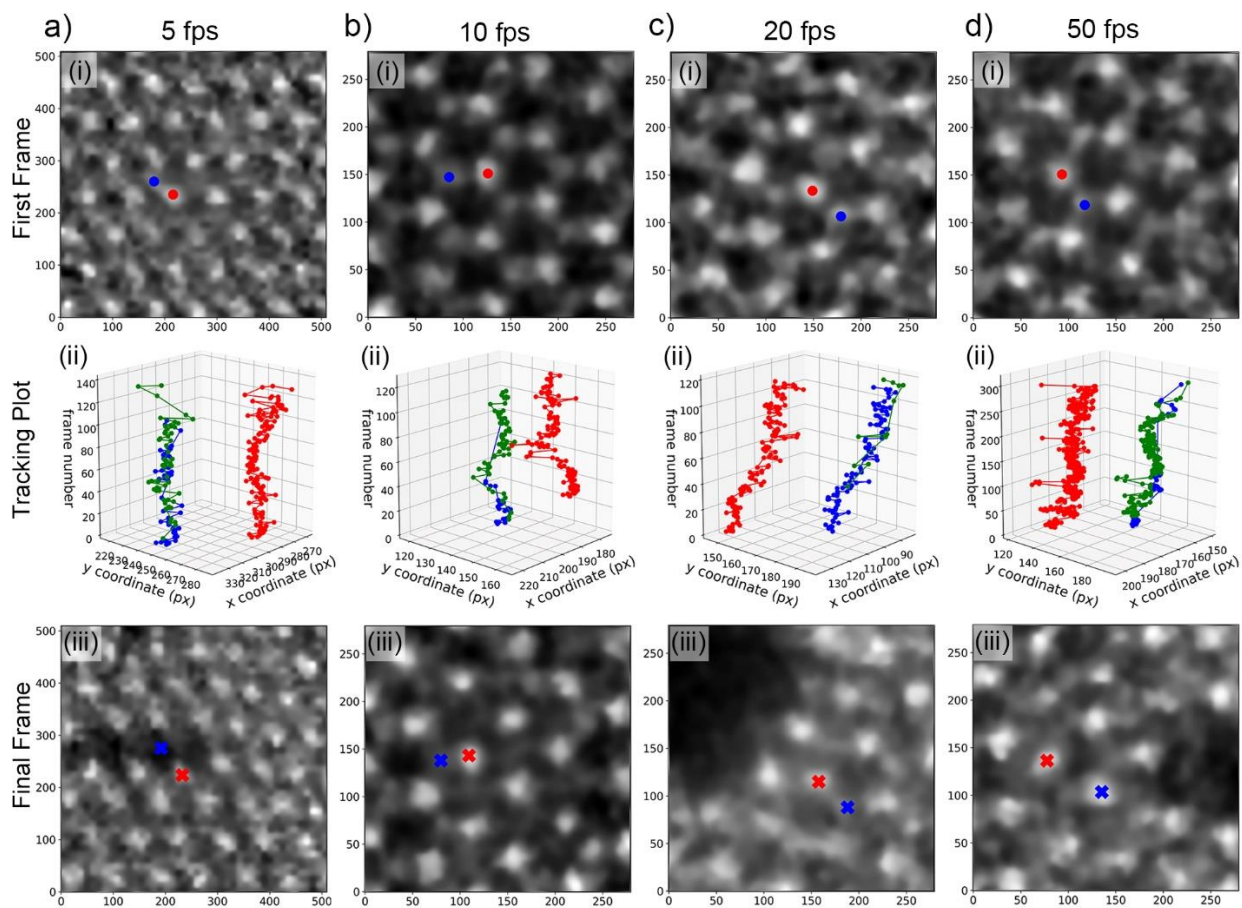

**Figure 4: (a-d)** Analysis from 5, 10, 20 and 50 fps acquired milling processes. *(i)* The initial processed frame with the selected Mo and S site selected to be tracked. *(ii)* A three dimensional plot that tracks the location and movement and change in the class of the atomic site with red = Mo, blue = S2, and green = S1 site or single $V_S$. *(iii)* The final processed frame showing the location of the atomic species or where the site was removed. Image FOV in (a) 3.0x3.0 nm and (b-d) 1.6x1.6 nm

Full frames of the tracking can be plotted and trends in the data can be clearly seen to a greater degree than single atomic columns. Highlighted here are the example 10 and 50 fps datasets in Figure 5. In the 10 fps the shift in the later frames as the lattice is compressed on the right side of the FOV can be seen as the SVL grows before the formation of a line defect neighboring the SVL. In the top view of the 50 fps scan, one can observe the atomic disordering near the bottom of the field of view as a nanopore nucleates and then expands. Throughout the reaction the nearest neighbor distance of each location can be tracked demonstrating the capability to track atomic disordering and lattice distortion of the structure and directly tie these changes to the dosage applied to the system. Additionally, when observing the global data, as expected, it was found that the percentage of Mo and S that left the field of view prior to the formation of a nanopore was independent of the speed of acquisition. To form the nanopore the number of Mo decreased by a percentage of 7±9% while a lot more S left the system at 51±9%. The small decrease and large error associated with the Mo atom count can be attributed to the atomic disordering and lattice shift that occur as a nanopore forms. When considering the decrease in S, this much larger percentage of S removal is to be expected as S is preferentially removed from the MoS$_2$ lattice during electron beam irradiation. Additionally, this percentage makes theoretical sense as, after nanopore formation, a common edge reconstruction in the MoS$_2$ system is the formation of the Mo$_6$S$_6$ nanowire edge structure. Decreasing the local quantity of S by roughly 50% satisfies the local stoichiometry needed to facilitate the formation of this structure.

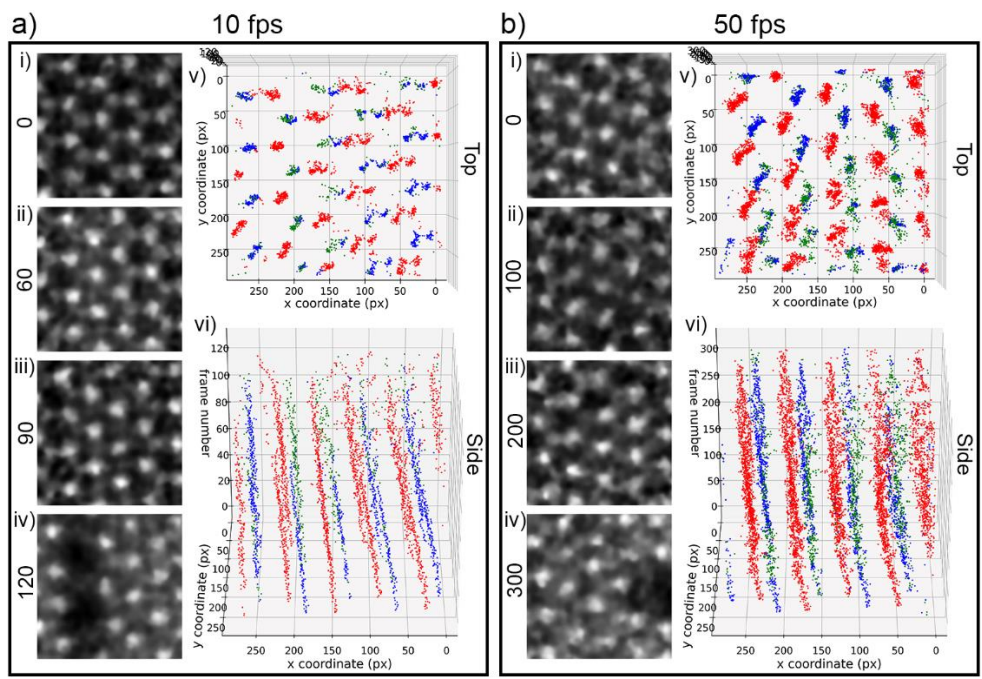

**Figure 5: (a-b)** Tracking of the full frames (i-iv) of different scans taken at 10 and 50 fps respectively going from pristine lattice (i) to the disordering and formation of defects (ii-iii) until the formation of the nanopore (iv). The decoded tracking of the full frame of the atom sites, viewed from both the top (v) and side (vi) view to see the atomic motion in the xy plane as well as in time.

**Conclusions:**

In summary, this study demonstrates the successful enhancement of temporal resolution in STEM imaging without direct modifications to the electron microscope optics. By employing custom scan control with image processing techniques and implementing the ELIT workflow to develop a DCNN, a significant increase in the speed of determination of atomic position and species information at high speeds in STEM imaging was achieved. This approach facilitated a more comprehensive investigation of the kinetics involved in the defect formation process. The application of DCNN-enabled decoding of low-quality, high-temporal-resolution data provided valuable insights into the atomic disordering and reordering process. Specifically, the experiments focused on the MoS$_2$ materials system, utilizing a beam control system capable of data acquisition at scanning rates of up to 100 fps, resulting in small datasets with resolutions as low as 64x64 pixels. Through image processing, atomic resolution was attained, and the ELIT workflow successfully decoded the data into Mo, S2, and S1 classes, enabling accurate tracking of these elements throughout the milling process. This workflow enables the extraction of more useful information from small datasets without compromising temporal or spatial resolution. The integration of machine learning and beam control techniques in this manner holds the potential for the development of different electron beam control methodologies and a deeper understanding of early defect formation and hole nucleation and growth and edge state formation in 2D materials. These findings have implications for advancing atomic fabrication processes and fostering autonomous atomic-level control.

**Methods:**

Sample preparation:

Chemical vapor deposition was used to synthesize the $MoS_2$ monolayers at a growth temperature of 750 °C using a hinged tube furnace (Lindberg/Blue) equipped with a 2 in. diameter quartz tube. $SiO_2$ (300 nm)/Si was used as a growth substrate. These substrates were first cleaned with isopropanol and then spin-coated with perylene-3,4,9,10- tetracarboxylic acid tetrapotassium salt before being dried and placed face-down above an alumina crucible containing $MoO_3$ powder (~5 mg) (99.9%, Sigma-Aldrich). This crucible was then inserted into the center of the quartz tube. Another crucible was filled with S powder (~0.7 g) (99.998%, Sigma-Aldrich) and inserted at a region located 20 cm upstream from the other crucible in the quartz tube. When the center of the tube reaches 750 °C center of the tube, the temperature at the S powder crucible was around 180 °C. The tube was then evacuated to ~5 mTorr. At this point a 70 sccm (standard cubic centimeters per minute) flow of Ar at atmospheric pressure was established. The furnace was heated to 750 °C with a ramping rate of 30 °C/min and held at this temperature for 4–6 min. The furnace is allowed to naturally cool down to room temperature.

These $MoS_2$ monolayers are then transferred from the $SiO_2$/Si substrate onto gold mesh holey carbon Quantifoil TEM grids using a wet transfer process for STEM imaging. PMMA (Poly(methyl methacrylate)) was spun coated onto the $SiO_2$/Si substrate with monolayer $MoS_2$ crystals on the surface at 500 rpm for 15 s followed by 3000 rpm for 50 s. The PMMA-coated sample was then placed into a 1 M KOH solution to remove the $SiO_2$ substrate. This leaves the PMMA film with the $MoS_2$ monolayer on the solution surface. The KOH residue was then rinsed from the sample film with DI water. A TEM grid was then placed into a funnel filled with DI water. Using a glass slide the washed PMMA/$MoS_2$ film was then transferred to float on the DI water. As the water drains the film is then placed onto the TEM grid. The samples are then baked to dry at 80 °C before being placed into acetone to soak for 12 hours to remove remaining PMMA before a final rinse in IPA, leaving a clean monolayer surface. Prior to STEM experiments, the specimens were baked at 160°C in vacuum overnight to reduce surface contamination.

STEM imaging:

An aberration corrected Nion UltraSTEM100 electron microscope were used for all the STEM imaging experiments while operating at 60kV. The custom-built scanning feedback control system described in this study was based on a National Instruments DAQ (PXIe-6124) and field-programmable gate array (FPGA) system (PXIe-7856R) in a PXIe-1073 chassis that was interfaced with the beam control system of the Nion UltraSTEM100 for spiral scans. Input coordinates from customizable MatLab code were then put into a LabView program that was used to control the STEM scan coils to position the focused STEM probe along well-defined scanning pathway shapes. The custom scan control system has a maximum IO rate of 2M Samples/s.

**Author Contributions:**

MGB conducted STEM experiments, constructed DCNN and performed data analysis and was the primary contributor in writing the manuscript. SM, KMR and AG provided assistance and guidance with the coding and data analysis with additional support in the construction of the DCNN from AG, MZ and SVK. SJ provided experimental assistance with the beam control system during STEM experiments. RRU provided guidance during experiment planning. All authors additionally contributed and approved the final manuscript.

**Acknowledgements:**

All STEM experimental work was performed at the Center for Nanophase Materials Sciences (CNMS), which is a US Department of Energy, Office of Science User Facility at Oak Ridge National Laboratory. The synthesis of MoS$_2$ was performed by Kai Xiao and development of the scan control was supported by the U.S. Department of Energy, Office of Basic Energy Sciences, Division of Materials Sciences and Engineering.